\documentclass[10pt]{article}

\usepackage{amsmath}
\usepackage{amssymb}

\usepackage{graphicx}

\usepackage{cite}

\usepackage{color} 

\usepackage[labelfont=bf,labelsep=period,justification=raggedright]{caption}

\makeatletter
\renewcommand{\@biblabel}[1]{\quad#1.}
\makeatother

\date{}

\begin{document}

\def\be{\begin{equation}}
\def\brho{\boldsymbol{\rho}}
\def\bzeta{\boldsymbol{\zeta}}
\def\ee{\end{equation}}
\def\bc{\begin{center}} 
\def\ec{\end{center}}
\def\bea{\begin{eqnarray}}
\def\eea{\end{eqnarray}}
\newcommand{\avg}[1]{\langle{#1}\rangle}
\newcommand{\Avg}[1]{\left\langle{#1}\right\rangle}

% Title must be 150 characters or less
\begin{flushleft}
{\Large
\textbf{Social interaction, noise and antibiotic-mediated switches in the intestinal microbiota}
}
% Insert Author names, affiliations and corresponding author email.
\\
Vanni Bucci$^{1,3,\ast}$, 
Serena Bradde$^{1,3,\ast}$, 
Giulio Biroli$^{2}$,
Joao B. Xavier$^{1,\ast}$
\\
\bf{1} Program in Computational Biology, Memorial Sloan-Kettering Cancer Center, New York, U.S.A.
\\
\bf{2} Institut Physique Th\'eorique (IPhT) CEA Saclay, and CNRS URA, Gif Sur Yvette, France
\\
\bf{3} joint first authorship
\\
$\ast$ E-mail: bucciv@mskcc.org, serena.bradde@pasteur.fr, xavierj@mskcc.org
\end{flushleft}

% Please keep the abstract between 250 and 300 words
\section*{Abstract}
The intestinal microbiota plays important roles in digestion and resistance against entero-pathogens. As with other ecosystems, its species composition is resilient against small disturbances but strong perturbations such as antibiotics can affect the consortium dramatically. Antibiotic cessation does not necessarily restore pre-treatment conditions and disturbed microbiota are often susceptible to pathogen invasion. Here we propose a mathematical model to explain how antibiotic-mediated switches in the microbiota composition can result from simple social interactions between antibiotic-tolerant and antibiotic-sensitive bacterial groups. We build a two-species (e.g. two functional-groups) model and identify regions of domination by antibiotic-sensitive or antibiotic-tolerant bacteria, as well as a region of multistability where domination by either group is possible. Using a new framework that we derived from statistical physics, we calculate the duration of each microbiota composition state. This is shown to depend on the balance between random fluctuations in the bacterial densities and the strength of microbial interactions. The singular value decomposition of recent metagenomic data confirms our assumption of grouping microbes as antibiotic-tolerant or antibiotic-sensitive in response to a single antibiotic. Our methodology can be extended to multiple bacterial groups and thus it provides an ecological formalism to help interpret the present surge in microbiome data.
% Please keep the Author Summary between 150 and 200 words
% Use first person. PLoS ONE authors please skip this step. 
% Author Summary not valid for PLoS ONE submissions.   
\section*{Author Summary}
Recent applications of metagenomics have lead to a flood of novel studies and a renewed interest in the role of the gut microbiota in human health. We can now envision a time in the near future where analysis of microbiota composition can be used for diagnostics and the rational design of new therapeutics. However, most studies to date are exploratory and heavily data-driven, and therefore lack of mechanistic insights on the ecology governing these complex microbial ecosystems. In this study we propose a new model grounded on ecological and physical principles to explain intestinal microbiota dynamics in response to antibiotic treatment. Our model explains a hysteresis effect that results from the social interaction between two microbial groups, antibiotic-tolerant and antibiotic-sensitive bacteria, as well as the recovery allowed by stochastic fluctuations. We use singular value decomposition for the analysis of temporal metagenomic data, which supports the representation of the microbiota according to two main microbial groups. Our framework explains why microbiota composition can be difficult to recover after antibiotic treatment, thus solving a long-standing puzzle in microbiota biology with profound implications for human health. It therefore forms a conceptual bridge between experiments and theoretical works towards a mechanistic understanding of the gut microbiota. 

\section*{Introduction}
Recent advances in metagenomics provide an unprecedented opportunity to investigate the intestinal microbiota and its role in human health and disease \cite{Neish2009,Dethlefsen2007}. The analysis of microflora composition has a great potential in diagnostics \cite{Jones2011} and may lead to the rational design of new therapeutics that restore healthy microbial balance in patients \cite{Khoruts2011, Borody2004, Ruder2011}. Before the clinical translation of human microbiome biology is possible, we must seek to thoroughly understand the ecological processes governing microbiota composition dynamics and function.

The gastro-intestinal microbiota is a highly diverse bacterial community that performs an important digestive function and, at the same time, provides resistance against colonization by entero-pathogenic bacteria \cite{Pultz2005, Stecher2008, Endt2010}. Commensal bacteria resist pathogens thanks to resources competition \cite{Neish2009, Stecher2008}, growth inhibition due to short-chain fatty acid production \cite{Fukuda2011}, killing with bacteriocins \cite{Dabard2001,Corr2007} and immune responses stimulation \cite{Stecher2011,Keeney2011}. However, external challenges such as antibiotic therapies can harm the microbiota stability and make the host susceptible to pathogen  colonization \cite{Pamer2007, Dethlefsen2008, Willing2011,Ubeda2010,Bishara2008, Buffie2011}. 

Despite its importance to human health, the basic ecology of the intestinal microbiota remains unclear. A recent large-scale cross-sectional study proposed that the intestinal microbiota variation in humans is stratified and fits into distinct enterotypes, which may determine how individuals respond to diet or drug intake \cite{Arumugam2011}. Although there is an ongoing debate over the existence of discrete microbiome enterotypes \cite{Wu2011}, they could be explained by ecological theory as different states of an ecosystem \cite{Scheffer}.  Ecological theory can also explain how external factors, such as antibiotics, may lead to strong shifts in the microbial composition. A recent study that analyzed healthy adults undergoing consecutive administrations of the antibiotic ciprofloxacin, showed that the gut microbiota changes dramatically by losing key species and can take weeks to recover \cite{Dethlefsen2010}. Longitudinal studies, such as this one, suggest that many microbial groups can have large and seemingly random density variations in the time-scale of weeks \cite{Dethlefsen2006,Caporaso2011}. The observation of multiple microbial states and the high temporal variability highlight the need for ecological frameworks that account for basic microbial interactions, as well as random fluctuations \cite{Ley2006, Foster2008, Santos2010}.

Here we propose a possible model to study how the intestinal microbiota responds to treatment with a single antibiotic. Our model expands on established ecological models and uses a minimal representation with two microbial groups \cite{Mao-Jones2010} representing the antibiotic-sensitive and antibiotic-tolerant bacteria in the enteric consortium (Fig. 1). We propose a mechanism of direct interaction between the two bacterial groups that explains how domination by antibiotic-tolerants can persist even after antibiotic cessation. We then develop a new efficient framework that deals with non-conservative multi-stable field of forces and describes the role played by the noise in the process of recovery. We finally support our model by analyzing the temporal patterns of metagenomic data from the longitudinal study of Dethlefsen and Relman \cite{Dethlefsen2010}. We show that the dynamics of microbiota can be qualitatively captured by our model and that the two-group representation is suitable for microbiota challenged by a single antibiotic.
Our model can be extended to include multiple bacterial groups, which is necessary for a more general description of intestinal microbiota dynamics in response to multiple perturbations.

% Results and Discussion can be combined.

\section*{Results}

\subsection*{Mathematical model} 
\begin{figure}[t]
\begin{center}
\includegraphics[width=0.5\columnwidth]{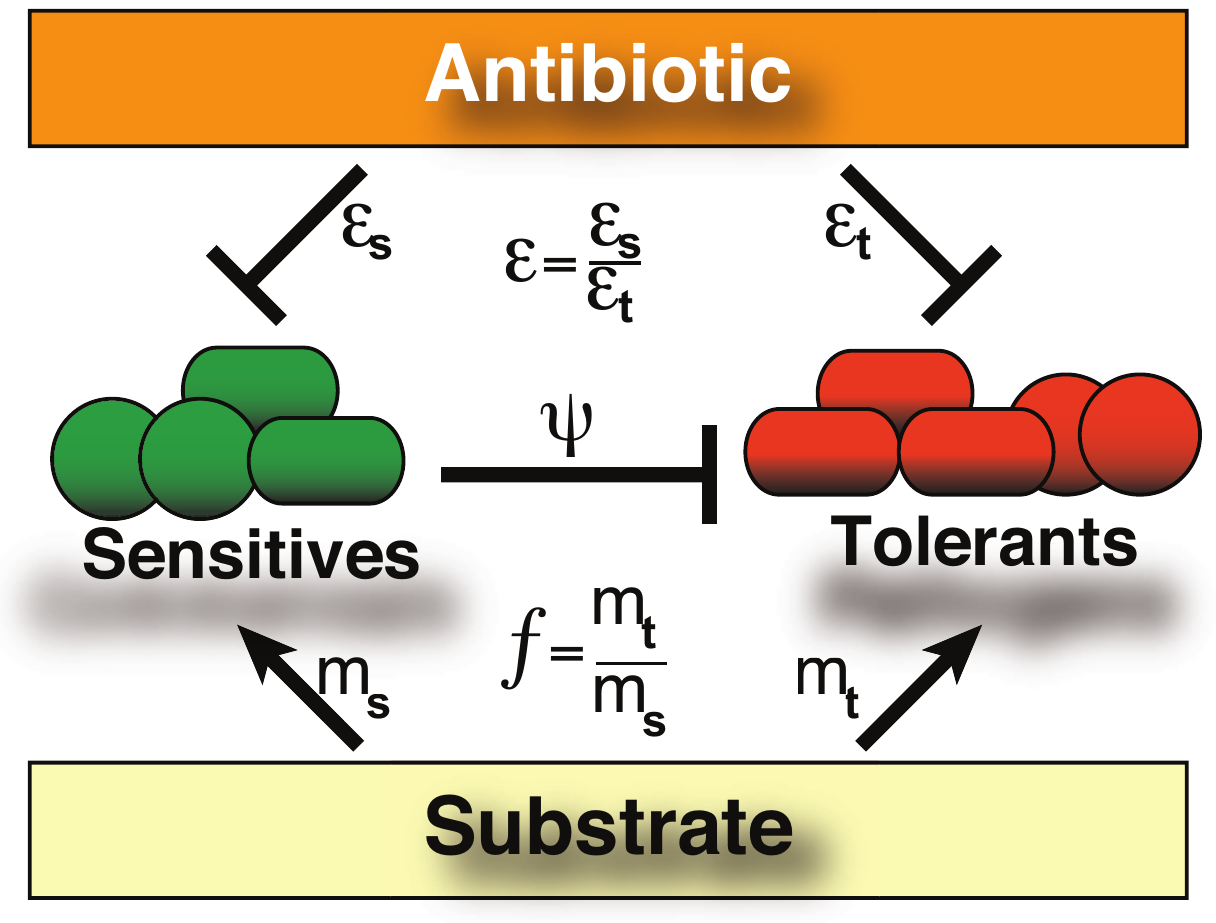}
\end{center}
\caption{The two-group model of the intestinal microbiota with antibiotic-sensitive and antibiotic-tolerant bacteria. Antibiotic sensitives can inhibit the growth of tolerants and both groups compete for the same growth substrate. Model parameters $\epsilon_s$ and $\epsilon_t$ represent the antibiotic sensitivity of sensitive and tolerant bacteria (where $\epsilon_s < \epsilon_t$), $m_s$ and $m_t$ represent their affinities to substrate and $\psi$ represents the inhibition of tolerants by sensitives.}
\label{fig:Scheme1}
\end{figure}
We model the microbiota as a homogeneous system where we neglect spatial variation of antibiotic-sensitive ($s$) and antibiotic-tolerant ($t$) bacterial densities. Their evolution is determined by growth on a substrate and death due to natural mortality, antibiotic killing and social pressure. 
With these assumptions, we introduce, as a mathematical model, two coupled stochastic differential equations for the  
density of sensitives and tolerants ($\rho_s$ and $\rho_t$) normalized with respect to the maximum achievable microbial density:
\begin{eqnarray} 
\frac{d\rho_s}{dt}&&=\frac {\rho_s}{\rho_s+f\rho_t}-\epsilon \rho_s+\xi_s (t)=F_s(\brho)+\xi_s(t) \label{eq:noise1}\\
\frac{d\rho_t}{dt}&&=\frac{f\rho_t}{\rho_s+f\rho_t}-\psi  \rho_s \rho_t-\rho_t+\xi_t(t)=F_t(\brho)+\xi_t(t)\label{eq:noise2}
\end{eqnarray}   
In the physics literature these types of equations represent stochastic motion in a non-conservative force field ${\bf F}$.
The first terms in ${\bf F}$ correspond to the saturation growth terms representing the indirect competition for substrate and depend on $f$, which is the ratio of the maximum specific growth rates between the two groups. If $f >1$ tolerants grow better than sensitives on the available substrate and the reverse is true for $f <1$.  They effectively describe a microbial system with a growth substrate modeled as a Monod kinetic \cite{Monod1949} in the limit of quasi-steady state approximation for substrate and complete consumption from the microbes (see Methods for details). Both groups die with different susceptibility in response to the antibiotic treatment, which is assumed to be at steady-state. $\epsilon$ defines the ratio of the combined effect of antibiotic killing and natural mortality rates between the two groups (see Methods for details). While the system can be studied in its full generality for different choices of $\epsilon$, we consider the case of $\epsilon >1$ because it represents the more relevant case where sensitives are more susceptible to die than tolerants in the presence of the antibiotic. A possible  $\epsilon(t)$ that mimics the antibiotic treatments is a pulse function. With this, we are able to reproduce realistic patterns of relative raise (fall) and fall (raise) of sensitives (tolerants) due to antibiotic treatment as we show in Fig. S4 in the Supplementary Information (SI) Text. Additionally, we introduce the social interaction term between the two groups, $\psi\rho_t\rho_s$, to implement competitive growth inhibition \cite{Stecher2011,Xavier2011}. In particular, we are interested in the case where the sensitives can inhibit the growth of the tolerants ($\psi>0$), which typically occurs through bacteriocin production \cite{Bucci2011}. Finally we add a stochastic term $\boldsymbol{\xi}$ that models the effect of random fluctuations (noise), such as random microbial exposure, which we assume to be additive and Gaussian. The analysis can be generalized to other forms of noise such as multiplicative and coloured. 

\subsection*{Antibiotic therapy produces multistability and hysteresis}

We first analyzed the model in the limit of zero noise, $\boldsymbol{\xi}=0$. In this case, we were interested in studying the steady state solutions that correspond to the fixed-points of equations (\ref{eq:noise1},\ref{eq:noise2}) and are obtained imposing ${\bf F}=0$.  We found three qualitatively-distinct biologically meaningful states corresponding to sensitive domination, tolerant domination and sensitive-tolerant coexistence (see SI Text). We evaluated the stability of each fixed point (see SI Text) and identified three regions within the parameter space (Fig.~2A). In the first region the effect of antibiotics on sensitive bacteria is very low ($f\epsilon <1$) and domination by sensitives is the only stable state (sensitives monostability). In the second region the effect of the antibiotic on sensitives is stronger than their inhibition over tolerants ($f\epsilon>1+\psi/\epsilon$) and the only stable state is domination by tolerants (tolerants monostability). Finally, in the third region ($1<f\epsilon<1+\psi/\epsilon$) both sensitive and tolerant dominations are possible and stable, while the third coexistence fixed point is unstable (bistability) (see SI Text). This simple analysis shows that multistability can occur in a gut microbiota challenged by an antibiotic where one group directly inhibits the other (i.e. through the $\psi$ term). Furthermore, it suggests that multistability is a general phenomenon since it requires only that antibiotic-sensitive and antibiotic-tolerant bacteria have similar affinities to nutrients. This is a realistic scenario because tolerants, such as vancomycin resistant \emph{Enterococcus} \cite{Ubeda2010}, are often closely related to other commensal but antibiotic-sensitive strains and therefore should have similar affinity to nutrients. 
\begin{figure}[!h]
\begin{center}
\includegraphics[width=0.6\columnwidth]{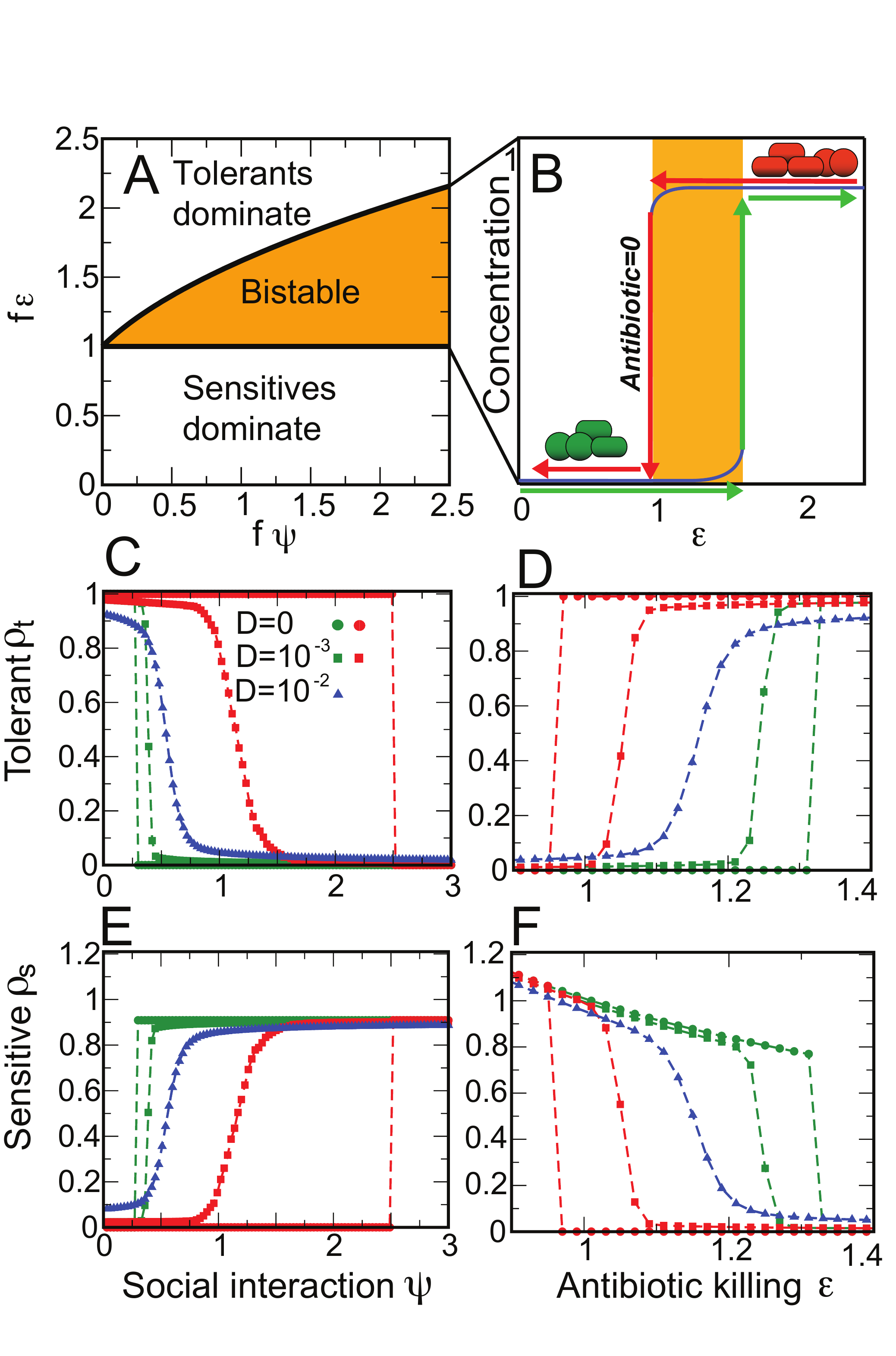}
\end{center}
\caption{Multistability and hysteresis in a simple model of the intestinal microbiota. A: phase diagram showing the three possible stability regions. Antibiotic-sensitive bacteria dominate when $f \epsilon <1$ and antibiotic-tolerant bacteria dominate when 
$f\epsilon> 1/2+\sqrt{1+4f\psi/2}$ and therefore these are regions of monostability. There is a region of bistability between the two regions where domination by either sensitives or tolerants is possible. B: schematic display of the hysteresis phenomenon explaining cases where antibiotic treatment produces altered microbiota (i.e. tolerants domination) that persists long after antibiotic cessation. C-F: mean density values obtained simulating the Langevin dynamics for a maximum time $T = 10000$ after an instantaneous change of the  parameter $\psi$ (C and D) and $\epsilon$ (E and F). These averages are obtained over $1000$ noise realizations. C, D and E, F show the antibiotic-tolerants or antibiotic-sensitives densities, respectively, as a function of the social interaction parameter ($\psi$) with $f\epsilon=1.21$ or the antibiotic killing ($\epsilon$) with $f\psi=0.77$.} 
\label{fig:stability}
\end{figure}
Finally, 
the solution of equations (\ref{eq:noise1}) and (\ref{eq:noise2}) reveals that hysteresis is present for values of $f\epsilon$ and$f\psi$  
leading to multistability (Fig. 2B). Similarly to magnetic tapes, such as cassette or video tapes, which remain magnetized even after the external magnetic field is removed (i.e. stopping the recording), a transient dose of antibiotics can cause a microbiota switch that persists for long time even after antibiotic cessation.

\subsection*{Noise alters stability points} 
The previous analysis shows the existence of multistability in the absence of noise. However, the influx of microbes from the environment and/or the intra-population heterogeneity are expected in realistic scenarios and affect the bacterial density evolution in a non-deterministic fashion. This raises the question of how the noise alters the deterministic stable states and their stability criteria. We assume that the noise is a fraction of bacteria $\boldsymbol{\xi}(t)$ that can be added (or removed) at each time step, but on average has no effect since $\langle \boldsymbol{\xi}\rangle=0$. This assumption is justified by the fact that a persistent net flux of non-culturable bacteria from the environment is unrealizable. We also assume that this random event at time $t$ is not correlated to any previous time $t'$, which corresponds to $\langle \xi_k(t) \xi_{k'}(t')\rangle=D\delta(t-t')\delta_{kk'}$, where $D$ characterizes the noise amplitude and $\delta$ is the Dirac delta function. We calculated the stationary probability of the microbiota being at a given state by solving the stationary Fokker-Planck Equation (FPE) \cite{Gardiner1997} corresponding to the Langevin equations (1,2):
\begin{equation}\label{eq:FPE2} 
- \boldsymbol{\nabla} \cdot(\mathbf{ F} \;P_s) +\frac{D}{2}\nabla^{2}P_s=0.
\end{equation}
By numerically solving equation (\ref{eq:FPE2})  as described in \cite{Galan2007}, for increasing $D$, we find that for small values of $D$ the most probable states coincide with the deterministic stable states given by ${\bf F}=0$ (Fig. 3A). However, by increasing $D$ the distribution $P_s$ spreads and the locations of the most probable states change and approach each other. As a consequence, the probability of an unstable coexistence, characterized by $\rho_s>0$ and $\rho_t>0$, increases thus avoiding extinction. This intuitively justifies how recovery to a sensitive-dominated state within a finite time after antibiotic cessation becomes possible with the addition of the noise. Without noise, the complete extinction of sensitive bacteria would have prevented any possible recolonization of the intestine. Beyond a critical noise level ($D_c$) bistability is entirely lost and the probability distribution becomes single-peaked with both bacterial groups coexisting. 
The microbiota composition at the coexistence state can be numerically determined from the solution of $P_s(\rho_s,\rho_t)$, as shown in Fig. 3B and Supplementary Video S1. Further investigations based on analytical expansion of the Langevin equations (see Methods) show that for small random fluctuations, $D\ll D_c$, the first noise-induced corrections to the deterministic density are linearly dependent on $D$ with a proportionality coefficient determined by the nature of the interactions (insets in Fig. 3B). These linear correction terms can be obtained as a function of the model parameters and, after substituting a particular set of values in the bistable region ($f=1.1, \epsilon= 1.1$ and $\psi=0.4$), they are $\langle\zeta^{(1)}_s\rangle=-4.3 \;D$ for sensitives and $\langle\zeta^{(1)}_t\rangle=4.4\;D$ for tolerants.
These numbers are different from those reported in the insets of Fig. 3B. However the discrepancy is due to the propagation of the boundary conditions when numerically solving the solution of the FPE using finite elements (see SI Text). 

\begin{figure}[!ht]
\begin{center}
\includegraphics[width=0.6\columnwidth]{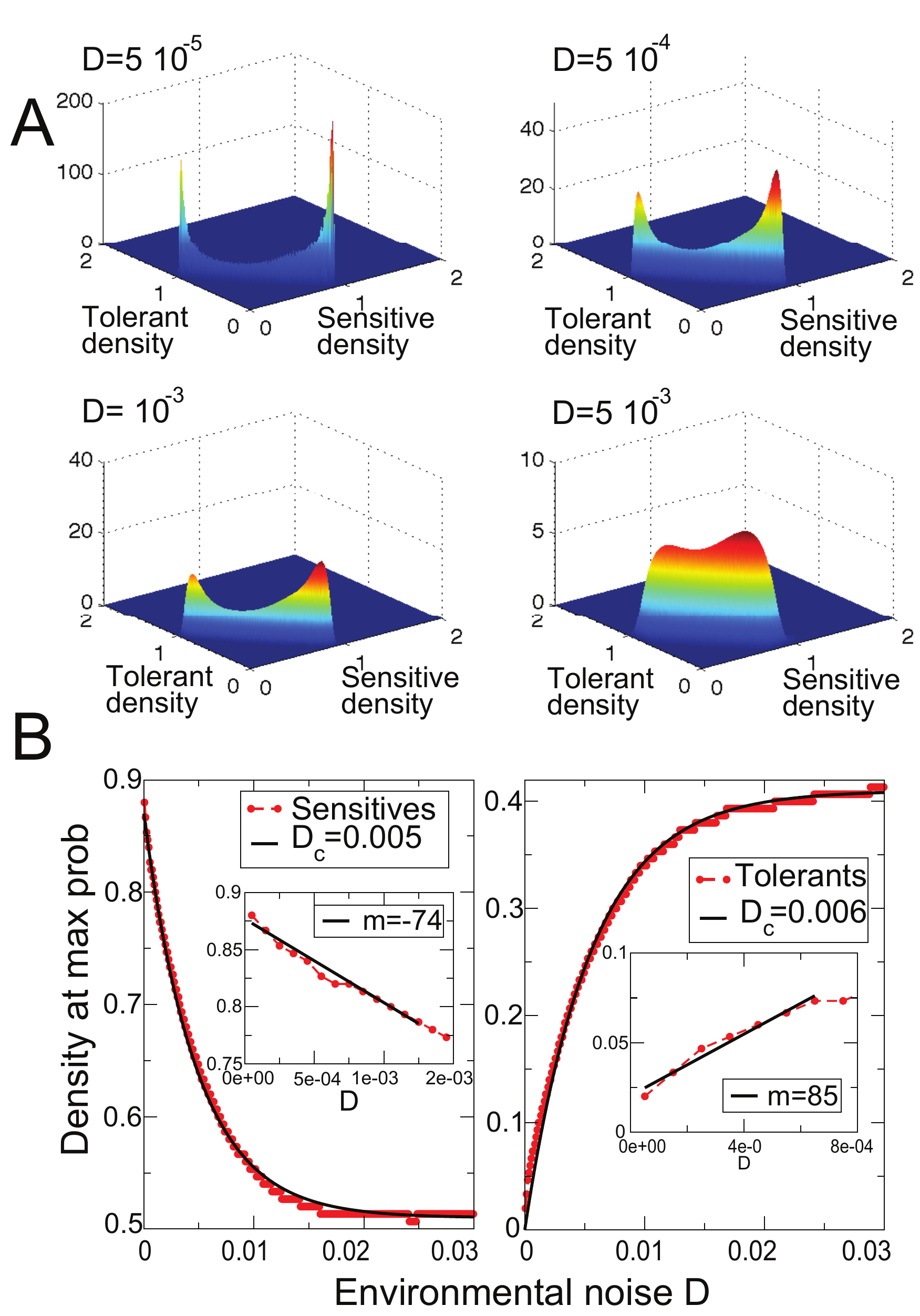}
\end{center}
\caption {Most probable microbiota states change from bistable scenario to mono-stable coexistence with increasing noise. A: the bacterial density joint probability distribution  determined by solving the Fokker-Planck equation (\ref{eq:FPE2}) for four different values of the environmental noise. B: the bacterial densities at the peaks of $P(\rho_s,\rho_t)$ as a function of the noise parameter $D$. Red symbols are data from the numeric solution of the Fokker-Planck equation and the black solid lines are the exponential fit. Parameters used: $f=1.1,\epsilon=1.1$ and $\psi=0.4$. The insets detail the linear regime.}
\label{fig:FPE_2}
\end{figure}

This has important biological implications since it suggests that extinction is prevented and, more importantly, that a minority of environmental microbes can settle in the gut at a rate that depends on the strength of their social interaction with the established microbiota. 

The introduction of random perturbation affects the stability criteria of the stable states.
In particular, we observe that the bistability region decreases when the noise amplitude $D$ increases (Fig. 2C-F). At the limit, when $D>D_c$ the bistability is entirely lost and the only stable state is the one where both groups coexist. This concept was previously hypothesized  but not explicitly demonstrated in a model of microbial symbionts in corals \cite{Mao-Jones2010}.

\subsection*{Noise affects the recovery time} 
Our model predicts that in absence of stochastic fluctuations the recovery time is larger than any observational time-scale so that it is impossible to revert to the conditions preceding antibiotic perturbation (see Fig. S4 in SI Text).  In reality, data show that this time can be finite and depends on the microbiota composition and the degree of isolation of the individuals \cite{Ubeda2010, Dethlefsen2010, Littman2011}. Thus, we aim to quantitatively characterize how the relative contribution of social interaction and noise level affects the computation of the mean residence time.  

In order to determine the relative time spent in each domination state, we compute the probability of residence $\pi_i(t)$ in each stable state $i={1,2,..,N}$ using master equations \cite{Gardiner1997}. This method is more efficient than simulating the system time evolution by direct integration of the Langevin equations because it boils down to solving a deterministic second-order differential equation. Furthermore, this approach scales up well when the number of microbial groups increases, in contrast to the numerical solution of the FPE which can become prohibitive when $N>3$. In our model, the master equations for the probability $\pi_i(t)$ of residing in the tolerant $i=1$ or sensitive $i=2$ domination state are: 
\begin{eqnarray}
\label{eq:residence}
\frac{d\pi_1(t)}{dt}=-\mathcal{P}_{1\to 2}\pi_1(t)+\mathcal{P}_{2\to 1}\pi_2(t) \nonumber \\
\frac{d\pi_2(t)}{dt}=-\mathcal{P}_{2\to 1}\pi_2(t)+\mathcal{P}_{1\to 2}\pi_1(t)
\end{eqnarray}
where $\mathcal{P}_{i\to j}$ is the transition rate from state $i$ to $j$, which  
can be obtained in terms of the sum over all the state space trajectories connecting $i$ to $j$. 

\begin{figure}[!ht]
\begin{center}
\includegraphics[width=0.6\columnwidth]{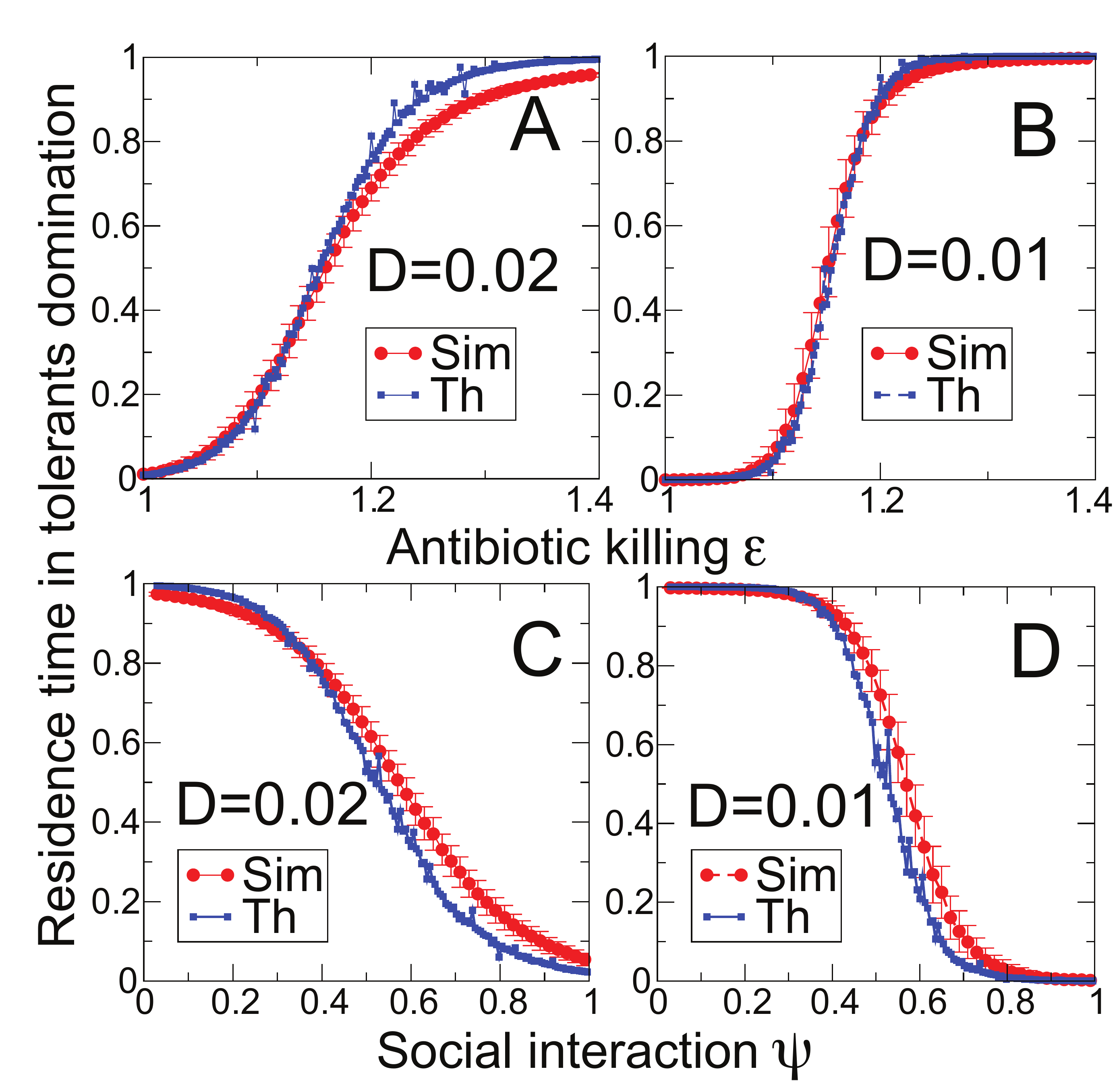}
\end{center}
\caption{Microbiota resident time in antibiotic-tolerant domination in function of the antibiotic action ($\epsilon$) and the social interaction ($\psi$) parameters. Blue circles show the theoretical predictions obtained by determining the probability of the most probable path. Red circles are obtained by simulating the Langevin dynamics over $10000$ iterations and averaged for $1000$ noise realizations. Higher order-corrections can be included to increase the theoretical estimation accuracy.}
\label{fig:time}
\end{figure}

By solving this system of equations at steady-state, we obtain the residence probabilities $\pi_1= \left(1+\mathcal{P}_{1\to 2}/\mathcal{P}_{2\to 1}\right)^{-1}$ and $ \pi_2=\left(1+\mathcal{P}_{2\to 1}/\mathcal{P}_{1\to 2}\right)^{-1}$. After computing the transition rate $\mathcal{P}_{i\to j} \propto e^{-\frac{ \mathcal{S(\brho^*)}}{D}}$ as a function of the parameters, as reported in the Methods, we determine $ \pi_2$, which is our theoretical prediction for the mean relative residence time $\langle t_2/(t_1+t_2) \rangle$ spent in the tolerant domination state (see Fig. 5). The theoretical predictions are in good agreement with those obtained by simulating the dynamics multiple times and averaging over different realizations of the noise. A first consequence from this analysis is that the time needed to naturally revert from the altered state depends exponentially on the noise amplitude ($1/D$). As such, we predict that for the case of an isolated system ($D\sim 0$) the switching time is exponentially larger than any other microscopic scale and the return to a previous unperturbed state is very unlikely. On the contrary, as the level of random exposure $D$ is increased, the time to recover to the pre-treated configuration decreases (see Fig. S4 in SI Text). Additionally, this method can be considered as a way to indirectly determine the strength of the ecological interactions between microbes which can be achieved by measuring the amount of time that the microbial population spends in one of the particular microbiota states. Therefore, it can potentially be applied to validate proposed models of ecological interactions by comparing residence times measured experimentally with theoretical predictions. 

\subsection*{Analysis of metagenomic data reveals antibiotic-tolerant and antibiotic-sensitive bacteria} 

We now focus on the dynamics of bacteria detected in the human intestine and test the suitability of our two-group representation by re-analyzing the time behaviour in the recently published metagenomic data of Dethlefsen and Relman \cite{Dethlefsen2010}. The data consisted of three individuals monitored over a 10 month period who were subjected to two courses of the antibiotic ciprofloxacin. Since the data are noisy and complex, and the individual subjects' responses to the antibiotic are distinct \cite{Dethlefsen2010}, identifying a time behaviour by manual screening is not a trivial task. We do it by using singular value decomposition (SVD) to classify each subject $p$ phylotype-by-sample data matrix $X^p$ into its principal components. 
Because of inter-individual variability we obtain, for each subject, the right and left eigenvectors associated to each eigenvalue. By ranking the phylotypes based on their correlation with the first two components we recover characteristic temporal patterns for each volunteer \cite{Alter2000, Brauer2006}. 

\begin{figure}[!ht]
\begin{center}
\includegraphics[width=0.8\textwidth]{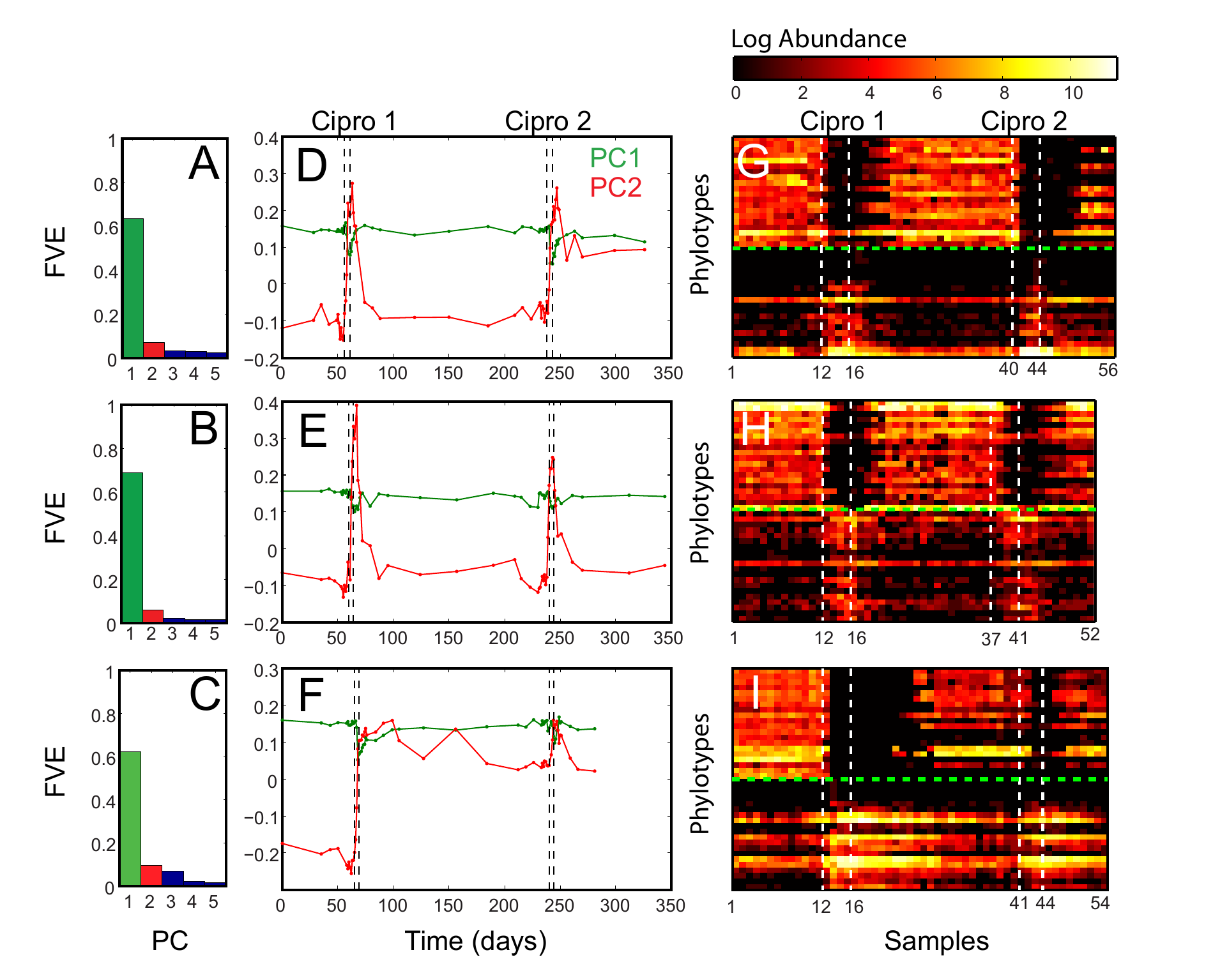}
\end{center}
\caption{Analysis of microbiota response to the antibiotic ciprofloxacin from three subjects \cite{Dethlefsen2010} using singular value decomposition identifies antibiotic-sensitive and antibiotic-tolerant bacteria. A-C: fraction of variance explained by the five most dominant components. D-F: plot of each sample component 1 (green) and 2 (red) coordinates versus sample time. G-I: sorting of the phylotypes log2-transformed abundance matrix based on the correlation within the two principal component. Above (below) the green dashed lines, we display the time series of the top 20 phylotypes strongly correlated (anti-correlated) with component 1 and anti-correlated (correlated) with 2 and dropping (increasing) during treatment, which we identify as sensitves (tolerants). Subject 3 (C,F,I) displays absence of sensitive bacteria for a prolonged period of about 50 days after the first antibiotic treatment. This confirms the fact that microbiota response to antibiotic can differ from subject to subject. Additionally, it also supports our model prediction of remaining locked in a tolerant-dominated state after antibiotic treatment cessation.}
\label{fig:data}	
\end{figure}

In all three subjects, we observe that, in spite of the individualized antibiotic effect, the two dominant eigenvalues or principal components together capture about 70\% of the variance observed in the data (Fig. 5A-C). Invariably, the first component shows a decrease in correspondence to antibiotic treatment and  reflects the behaviour of antibiotic-sensitive bacteria (green line in Fig. 5D-F). Conversely, the second component increases with the antibiotic treatment and represents antibiotic-tolerants (red line in Fig. 5D-F).
The observation that each subject's microbiota can be decomposed into two groups of bacteria with opposite responses to antibiotics supports the validity of the two-group approach used in our model. Classification of each individual's phylotypes as sensitive or tolerant can be obtained by determining their correlation with the two principal components (see SI Text) (information in the right-eigenarrays matrix from SVD). Bacteria correlated with component 1 are usually highly abundant before antibiotic treatment and drop strongly during treatment, often below detection. Vice-versa, bacteria correlated with component 2 are typically in low abundance before the antibiotic and increase with antibiotic administration (Fig. 5G-I). Interestingly, despite significant inter-individual differences in recovery time (Fig. 5G-I) and individualized response of each subject, the data show that in each individual the majority of bacteria are antibiotic-sensitive and only a small but significant fraction are tolerant to ciprofloxacin (see SI Text). 
The recognition of these time-patterns could be considered as a possible tool to indirectly determine the susceptibility of non-culturable commensal bacteria to FDA-approved antimicrobial compounds. However, the presence of strains in the same phylotypes that display both behaviors in response to the drug may constitute a significant challenge for the success of this method.

The time evolution of the phylotypes (Fig. 5G-I)  qualitatively agrees with our theoretical prediction that after the antibiotic administration the system moves fast, meaning in a time smaller than any other observable time-scale, into a new stable state with less sensitives and more tolerants. Further, the data also suggest that the return to sensitive domination happens after a recovery-time scale that depends on the microbial composition. 

\section*{Discussion}
We present a model  of inter-bacterial interactions that explains the effect of antibiotics and the counter-intuitive observation that an antibiotic-induced shift in microbiota composition can persist even after antibiotic cessation. Our analysis predicts a crucial dependence of the recovery time on the level of noise, as suggested by experiments with mice where the recovery depends on the exposure to mice with untreated microbiota \cite{Ubeda2010}. 
The simple model here introduced is inspired by classical ecological modeling such as competitive Lotka-Volterra models \cite{Sole2006,Zhu2009}, but relies on mechanistic rather than phenomenological assumptions, such as the logistic growth. Although more sophisticated multi-species models include explicit spatial structure to describe microbial consortia \cite{Mitri2011, Bucci2011, Munoz2010, Munoz2011}, our model is a first attempt to quantitatively analyze the interplay between microbial social interactions ($\psi$) and stochastic fluctuations ($D>0$) in the gut microbiota.  We find that these two mechanisms are the key ingredients to reproduce the main features of the dynamics in response to antibiotic (sudden shifts and recovery). Our model can be easily generalized to include spatial variability and more complicated types of noise. Therefore we provide a theoretical framework to quantify microbiota resilience against disturbances, which is an importance feature in all ecosystems \cite{Holling1973}. By introducing a new stochastic formulation, we were able to characterize composition switches within the context of state transition theory \cite{Langer1967,Langer1968}, an important development over similar ecological models of microbial populations  \cite{Mao-Jones2010}. We present a new method to calculate the rate of switching between states that identifies the most likely trajectory between two stable states and their relative residence time, which can be subjected to experimental validation. Finally, we apply SVD to previously published metagenomic data \cite{Dethlefsen2010}, which allows us to classify the bacteria of each subject in two groups according to their temporal response to a single antibiotic. The SVD method has been used before to find patterns in temporal high-throughput data, including transcription microarrays \cite{Alter2000} and metabolomics \cite{Yuan2009}. Although our approach seems to capture well the main temporal microbiota patterns, we should note that the use of the Euclidean distance as a metric for microbiome analysis presents limitations and recent studies have proposed alternative choices \cite{Gonzalez2011,Hamady2009,Kuczynski2010}. We also opt for an indirect gradient analysis method \cite{Braak2004} because we are interested in emergent patterns from the data regardless of the measurements of the external environmental variable (i.e. presence or absence of the antibiotic) \cite{Kuczynski2010}. 

We propose a mechanism of interaction between two bacterial groups to explain the lack of recovery observed in the experiments that can be validated in the near future. Although training the model with the available data sets would be of great interest, this will not be useful in practice because we need more statistical power to be predictive.
However, we anticipate that a properly validated mathematical model of the intestinal microbiota will be a valuable tool to assist in the rational design of antibiotic therapies. For example, we predict that the rate of antibiotic dosage will play a crucial role. In order to let the microbiota recover from antibiotic treatment, it is better to gradually decrease antibiotic dosage at the first sign of average microbiota composition change, which has to be larger than the threshold community change represented by the day-to-day variability \cite{Caporaso2011}, rather than waiting for tolerant-domination and then stopping antibiotic treatment. 

We show here the application of our theory to a two-bacterial group scenario because we are interested in the microbiota response when challenged with a single antibiotic. However, in more realistic conditions the microbiota is subjected to different types of perturbations, which may drive it towards more alternative stable states. Our theory of the microbial-states switches characterization can be naturally extended to more than two states and consists of the solution of the linear system of equations $\boldsymbol{\pi}{\bf P}=0$, where $\boldsymbol{\pi}$ is the array of probability of residing in each stable state and ${\bf P}$ is the matrix of transition rates among the states.

The ongoing efforts to characterize the microbial consortia of the human microbiome can yield tremendous benefits to human health \cite{Turnbaugh2009, Ichinohe2011, Veiga2010, Lee2010}. Within the next few years, we are certain to witness important breakthroughs, including an increase in the number of microbiomes sequenced as well as in sequencing depth. Yet, without the proper ecological framework these complex ecosystems will remain poorly understood. Our study shows that, as in other complex microbial ecosystems, ecological models can be valuable tools to interpret the dynamics in the intestinal microbiota.

% You may title this section "Methods" or "Models". 
% "Models" is not a valid title for PLoS ONE authors. However, PLoS ONE
% authors may use "Analysis" 

\section*{Methods}

\subsection*{Full model and simplification}
The model introduced in equations \ref{eq:noise1} and \ref{eq:noise2} is derived from the more detailed model described below. 
We model the bacterial competition in a well-mixed system in the presence of antibiotic treatment by means of the following stochastic differential equations:
\begin{eqnarray}\label{eqn:4dim}
&&\frac{dS}{dt}=K(S_0-S)-\frac{m_s S \rho_s}{B_s (S+a)}-\frac{m_t S \rho_t}{B_{t}(S+a)} \nonumber\\
&&\frac{d\rho_s}{dt}=\frac{m_s S \rho_s}{S+a}-\gamma A \rho_s - K \rho_s+\xi_{s}(t) \nonumber\\
&&\frac{d\rho_t}{dt}=\frac{m_t S \rho_t}{S+a}-\psi  \rho_s \rho_t - K \rho_t +\xi_{t}(t) \nonumber\\
&&\frac{dA}{dt}=K(A_0-A) 
\end{eqnarray}
where we account for two bacterial groups; the intestinal resident sensitive flora $\rho_s$ and an antibiotic tolerant one $\rho_t$. Additionally, we also consider the substrate $S$ and the antibiotic $A$ densities. The antibiotic time evolution is simply a balance between inflow and outflow (i.e. no decay due to microbial degradation) where $K$ is the system's dilution rate, which sets the characteristic microscopic time-scale, and $A_0$ is the constant density of the incoming antibiotic, which can be time dependent. Similarly the substrate concentration, $S$, results from a mass balance from influx and microbial consumption. As for the antibiotic, $S_0$ is the constant density of the incoming nutrient (i.e. the concentration of resources coming from the small-intestine).  The second and third terms in the right-hand side of the second equation in (\ref{eqn:4dim}) describe the amount of substrate consumed by bacterial growth assuming Monod kinetics where $m_s$ ($m_t$) is the maximum growth rate for sensitives (tolerants), $a$ is the half-saturation constant for growth, which parametrizes the bacterial affinity to the nutrient,  and $B_s$ ($B_t$) is the yield for growth for sensitives (tolerants). The last two equations describe how sensitives and tolerants grow on the substrate available and are diluted with the factor $K$. We mimic the effect of the antibiotic on the sensitives adding a term proportional to the sensitive density where the constant of proportionality $\gamma A$ is the antibiotic-killing rate. We also introduce a direct inhibition term $\psi\rho_s$, which mimics the inhibition of sensitive bacteria on the tolerants (social interaction). Finally the Gaussian random variables $\xi_{s}$, $\xi_{t}$ are the additive random patterns of exposure and represent the random microbial inflows (outflows) from (to) the external environment.

It is convenient to scale the variables and set the dilution rate to unity ($K=1$). Therefore, all the rates have to be compared with respect to the system characteristic dilution rate. Introducing 
$ \tilde{S} =S/S_0$ , $\tilde{\rho}_s =\rho_s/(B_{s} S_{0})$, $\tilde{\rho}_t =\rho_t/(B_{t} S_{0})$, $\tilde{A} =A/A_{0}$, $\tilde{\gamma}=(A_0 \gamma)/K $, $\tilde{\psi}=\psi/(K B_{s} S_{0}) $, $\tilde{m}_s=m_s/K$, $\tilde{m}_t=m_t/K $, $\tilde{a}=a/S_{0}$, $\tilde{\xi}_s=\xi_s/(B_{s} S_{0} K)$ and $\tilde{\xi}_t=\xi_t/(B_{t} S_{0} K)$ and dropping the tilde symbols, we obtain the following dimensionless model:
\begin{eqnarray}
\label{eq:adim}
&&\frac{dS}{dt}=1-S-\frac{m_{s}  \rho_s}{S+a}S-\frac{m_t\rho_t}{S+a} S \nonumber\\
&&\frac{d\rho_s}{dt}=\frac{m_{s} S}{S+a}\rho_s-{\gamma A \rho_s}-{\rho_s}+\xi_{s} \nonumber\\
&&\frac{d\rho_t}{dt}=\frac{m_{t} S }{S+a}\rho_t-{\psi  \rho_s \rho_t}-{\rho_t}+\xi_{t} \nonumber\\
&&\frac{dA}{dt}=1-A
\end{eqnarray}
If we assume that the antibiotic is a fast variable compared to the microbial densities ($\rho_s, \rho_t$) (i.e. the time-scale at which the antibiotic reaches stationary state is smaller than that of the bacteria), we can solve for $\frac{dA}{dt}=0$ and obtain $A=1$. 
If we also assume that the incoming substrate is all consumed in microbial growth, therefore maintaining the population in a stationary state with respect to the available resources, and that, similarly to the antibiotic, the resources equilibrate much faster than the bacterial densities (quasi-steady state assumption, $\frac{dS}{dt}=0$), we obtain that:
\begin{equation} \frac{S}{S+a}=\frac{1}{m_{s}\rho_s + m_{t}\rho_t}\,. 
\end{equation}
If we now define a new parameter $\epsilon=(\gamma+1)$ describing the relative ratio of the combination of antibiotic killing and natural mortality (i.e. wash-out) between sensitives and tolerants, the model reduces to the two variables model in $\boldsymbol{\rho}$ reported in equations (\ref{eq:noise1}-\ref{eq:noise2}).

\subsection*{Effective potential and location of long-term states}
The introduction of random noise has the important consequence of changing the composition of the stable states (Fig. 3A). In order to characterize this phenomenon, we expand the solution of the Langevin equations (\ref{eq:noise1}-\ref{eq:noise2}) around one of the stable states obtaining the following set of equations for the variable $\bzeta=\brho-\brho_i$:
\be
\frac{d \zeta_\iota}{dt}= \sum_\sigma \left.\frac{d F_\iota}{d \zeta_\sigma} \right|_{\brho_i}\zeta_\sigma + \frac{1}{2}\sum_{\sigma \kappa} \left.\frac{d F_\iota}{d \zeta_\sigma d\zeta_\kappa}\right|_{\brho_i} \zeta_\sigma \zeta_\kappa +\ldots+ \xi_\iota
\ee
where to simplify the notation we drop the explicit time-dependence. We can easily recognize the first derivative of the force on the right-hand side as the Jacobian matrix computed in one of the minima $ \left.\frac{d F_\iota}{d \zeta_\sigma} \right|_{\brho_i} = {\bf J}(\brho_i)$. This equation can be solved order by order by defining the expansion $\bzeta=\bzeta^{(0)} + \bzeta^{(1)}+\ldots$ and writing the equations for each order as:
\bea
\label{eqn:firstord}
\frac{d\zeta^{(0)}_\iota}{dt} & =& \sum_\sigma J_{\iota\sigma}(\brho_i) \zeta^{(0)}_\sigma + \xi_\iota \\
\label{eqn:secondord}
\frac{d\zeta^{(1)}_\iota}{dt} &=& \sum_\sigma  J_{\iota\sigma}(\brho_i)\zeta^{(1)}_\sigma + \frac{1}{2} \sum_{\sigma \kappa} V_{\iota\sigma\kappa}(\brho_i) \zeta^{(0)}_\sigma \zeta^{(0)}_\kappa\,.
\eea

Assuming that the initial condition at time zero is $\zeta_\iota(0)=0$, which can always be neglected for long-term behaviour, the solution of equation (\ref{eqn:firstord}) is
\be\label{eqn:solfirst}
\zeta^{(0)}_\iota(t)= \int_{0}^t dt'\,\sum_{\sigma} \left[ e^{\mathbf{J} (t-t')}\right]_{\iota\sigma} \xi_\sigma(t') .
\ee
This means that the average location of the minima at zero order is not modified by the noise since $\avg{\bzeta^{(0)}}\propto \avg{\boldsymbol{\xi}}=0$.
By computing the solution of the equation (\ref{eqn:secondord}) we similarly find that:
\be
\zeta^{(1)}_\iota(t)=\frac 1 2\int_0^t \sum_{\sigma\kappa\mu} \left[ e^{\mathbf{J} (t-t')}\right]_{\iota\sigma} V_{\sigma\kappa\mu} \;\zeta^{(0)}_\kappa(t')\; \zeta^{(0)}_\mu(t') dt'
\ee
The long-time average value of the first order correction now reads:
\be
\lim_{t\to\infty} \Avg{\zeta^{(1)}_\iota(t)}=\frac 1 2\lim_{t\to\infty} \int_0^t \sum_{\sigma\kappa\mu} \left[ e^{\mathbf{J} (t-t')}\right]_{\iota\sigma} V_{\sigma\kappa\mu} \Avg{\zeta^{(0)}_\kappa(t')\; \zeta^{(0)}_\mu(t')} dt'
\ee
The time integral can be easily computed assuming that the eigenvalues of $\mathbf{J}$ are negative, or at least their real part is, as it should be for stable fixed points; therefore we obtain that: 
 \be
\lim_{t\to\infty} \Avg{\zeta^{(1)}_\iota(t)}=\frac 1 2 \sum_{\sigma\kappa\mu} - \left[{\bf J}^{-1}\right]_{\iota\sigma} V_{\sigma\kappa\mu} \;\avg{\zeta^{(0)}_\kappa(\infty)\, \zeta^{(0)}_\mu(\infty)}\,.
\ee
Thus, we find that the effect of random fluctuations is to correct the value of the stable points as if an external field, proportional to strength of the fluctuations, was present. This field is equal to
the mean square displacement at large time opportunely weighted by the inverse of the curvature of the bare potential around the stable points, $\mathbf{J}(\brho_i)$. The correlation can be now computed using equation (\ref{eqn:solfirst}) and reads:	
\be
\avg{\zeta^{(0)}_\kappa(\infty)\, \zeta^{(0)}_\mu(\infty)}=\lim_{t\to\infty} \int_0^t dt'\int_0^t dt''\sum_{\sigma\sigma'}\left[ e^{\mathbf{J} (t-t')}\right]_{\kappa\sigma}\left[ e^{\mathbf{J} (t-t'')}\right]_{\mu\sigma'} \avg{\xi_\sigma(t')\xi_{\sigma'}(t'')}
\ee
Since $ \avg{\xi_\sigma(t')\xi_{\sigma'}(t'')}=D\delta_{\sigma\sigma'}\delta(t'-t'')$ the previous equation simplifies to 
\be
\avg{\zeta^{(0)}_\kappa(\infty)\, \zeta^{(0)}_\mu(\infty)}=\lim_{t\to\infty}  D\int_0^t dt'\sum_{\sigma}\left[ e^{\mathbf{J} (t-t')}\right]_{\kappa\sigma}\left[ e^{\mathbf{J} (t-t')}\right]_{\mu\sigma} \,.
\ee
which results in $\avg{\bzeta^{(1)}}\propto D$.

\subsection*{Theoretical estimate of the mean residence time}
The mean residence time in each state is proportional to the residence probability  $\pi_i(t)$ defined in equation (\ref{eq:residence}). To obtain it, we need to compute the transition rate $P_{i\to j}$ as a function of the model parameters as:
\begin{equation}
\label{eqn:rate}
\mathcal{P}_{i\to j}=\frac{1}{t_f-t_i}\int_{\brho_i}^{\brho_j} \mathcal{D}\brho\; P(\brho),
\end{equation}
where $t_i$ and $t_f$ are the initial and final time and $\mathcal{D}\brho$ is the functional integral over the trajectory $\brho(t)$. Each time trajectory $\brho(t)$, solution of equations (\ref{eq:noise1}-\ref{eq:noise2}), has an associated weight $P(\brho)$, defined as:
\be\label{eqn:weight}
P(\boldsymbol{\brho})=\int \mathcal{D} \boldsymbol{\xi} \; P(\boldsymbol{\xi}) \delta(\boldsymbol{\xi} - \dot{\brho} + {\bf F}(\brho)).
\ee
By discretizing the time so that $t=\ell\tau$ with $\ell=1,\ldots,M$ and $\tau$ the microscopic time step, we obtain that the Langevin equations can be written using the Ito prescription \cite{Gardiner1983} as:
\be\label{eqn:discr}
\frac{{\boldsymbol\rho}^\ell-{\boldsymbol\rho}^{\ell-1}}{\tau}={\bf F}({\boldsymbol\rho}^{\ell-1}) +{\boldsymbol\xi}^\ell
\ee
where we use the short notation $\brho(\ell\tau)=\brho^\ell$ and the initial value is $\brho^0=\brho_i$.
The time discretization allows us to interpret the functional integral in equation \eqref{eqn:weight} as:
\be 
P(\brho)=\int  \prod_{\ell=1}^{M}d\boldsymbol{\xi^\ell} \;P(\boldsymbol{\xi^\ell})\; \delta\left(
\brho^\ell-\brho^{\ell-1}-\left[{\bf F}({\boldsymbol\rho}^{\ell-1})+\boldsymbol{\xi}^\ell\right]\tau
\right)
\ee
Since the noise is Gaussian and white, its distribution now reads: 
\be\label{eqn:distr}
P\left(\boldsymbol{\xi}^\ell\right)=\left(\frac{\tau}{2\pi D}\right)^{1/2} e^{-\frac{\tau}{2 D} |\boldsymbol{\xi}^\ell|^2}.
\ee
This can be justified using the property of the delta-function $\int \delta(t-t') dt=1$ and its discrete time version  $\tau\sum_{i=1}^{M}f(\tau)\delta_{ij}=1$ so that $f(\tau)=\epsilon^{-1}$  follows and $\delta(t-t')\to\delta_{ij}/\tau$.

Using the properties of the delta function,
and integrating out all $\boldsymbol\xi^{\ell}$s, the continuous limit expression of equation (\ref{eqn:distr}) is
\begin{equation}
\label{eq:ProbTraj}
P(\boldsymbol{\rho}(t))=e^{-\frac{ \mathcal{S(\brho)}}{D}}
\end{equation}
where $S(\brho)=\frac{1}{2}\int_{t_i}^{t_f}  dt'\;|\boldsymbol{\dot{\rho}}(t') -  \mathbf{F}(\boldsymbol{\rho})|^2$ has an intuitive interpretation in thermodynamics and it is related to the entropy production rate \cite{Seifert2008}.
By using stationary-phase approximation, it turns out that in the computation of the rate defined in (\ref{eqn:rate}) only one path matters, $\brho^*$, which is the most probable path.
Higher order factors are proportional to the term $\Delta T= t_f-t_i$ \cite{Langer1967,Langer1968}, and therefore simplify with the denominator in equation (\ref{eqn:distr}). This comes from the fact that several almost optimal paths can be constructed starting from $\brho^*$. In the optimal path, the system stays in a stable state for a very long time, then it rapidly switches to the other stable state where it persists until $t_f$. By shifting the switching time one obtains sub-optimal paths that, at the leading order in $D$, give the same contribution of the optimal one and their number is directly proportional to $\Delta T$.
This leads to 
\be 
\mathcal{P}_{i\to j}({\boldsymbol\rho})\propto e^{-\frac{\mathcal{S}({\boldsymbol\rho}^*(t))}{D}} \int \mathcal{D} {\boldsymbol\rho} \;\exp\left(-\frac{1}{2D}\int dt dt'{\boldsymbol\rho}(t)\frac{ \delta^2\mathcal{S}({\boldsymbol\rho})}{\delta{\brho}(t) \delta{\brho}(t')}  \brho(t')\right).
\ee

The functional Gaussian integral can be computed \cite{Langer1967,Langer1968} and only provides a sub-leading correction to the saddle-point contribution resulting in the transition rate formula $\mathcal{P}_{i\to j} \propto e^{-\frac{ \mathcal{S(\brho^*)}}{D}}$, which is reported in the Results section.

We now need to determine the optimal path and its associated action $S(\brho^*)$. This path is defined as the one where the functional derivative of $S$ is set to zero such that the initial and final states are fixed. This produces a set of second-order differential equations
\be \label{secondorder}
\ddot{\rho}_\alpha=\sum_\beta F_\beta\frac{\partial F_\beta}{\partial \rho_\alpha}+\sum_\beta \dot{\rho}_\beta\left(\frac{\partial F_\alpha}{\partial \rho_\beta} - \frac{\partial F_\beta}{\partial \rho_\alpha}\right)\,
\ee
which can be solved imposing the initial conditions on $\brho_i$ and $\dot{\brho}(t_i)$.

It is easy to verify that the downhill solution is $\dot{\brho}=\bf{F}$ and it is associated with null action. Meanwhile, the ascending trajectory, which is the one  leading to a non-zero action and hence gives the transition rate value, is not given by  $\dot{\brho}=-{\bf F}$, as it would be for conservative field of forces. This means that in presence of a dissipative term the reverse optimal path from the minimum to the maximum is different with respect to the one connecting the maximum from the minimum of the landscape. 

As the last point, we want to show that the action associated to the optimal path can be further simplified by noticing that
\be
E=\frac{1}{2} \left(| \dot{\brho}|^2 - |\mathbf{F}(\brho)|^2\right)=0\,.
\ee
We can easily prove this condition by showing that the time derivative $dE/dt$ vanishes when equation (\ref{secondorder}) is satisfied and remembering that the optimal path connects two stable states where $\mathbf{F}=0$ and $\dot{\brho}=0$.  This property allows us to rewrite the action as:
\be
\mathcal{S}(\brho^*)=\int_{t_i}^{t_f}dt'\;\left[ |\dot{\brho}^*(t')|^2 - \dot{\brho}^*(t') \cdot \mathbf{F}(\brho^*(t'))\right]\,.
\ee

We solved numerically the equation \eqref{secondorder} using a trial-and-error approach. We varied the first-derivative at initial time in order to arrive as close as possible to the final point within some numerical precision. In principle the ideal trajectory connecting two stable points should be computed in the limit of $\dot{\brho}(t_i)\to 0$ but this trajectory will take infinite time.
We report three examples of most probable paths connecting the points $i$ to $j$ and reverse for a chosen set of $\dot{\brho}(t_i)$ in Fig.~S6 of the SI Text . 

\subsection*{Singular Value Decomposition}
We first rarefy the raw phylotypes counts matrix as in \cite{Dethlefsen2010}. We then normalize the logarithm of the counts according to the following procedure: 
1) we add one to all the phylotypes counts to take into account also for the non-detected phylotypes in each sample, 2) we log-transform the data and 3) we normalize the resulting matrix with respect 
to the samples averages. In formulae, the count associated to phylotype $i$ in sample $j$ for each subject $p$ is
$$X^p_{ij}= \log_2(Raw^p_{ij}+1)-\mu_j$$, 
where $\mu_j=\sum_{i=1}^N \log_2(Raw^p_{ij}+1)/N$ is the average value of the counts in each sample and $N$ is the total number of phylotypes.
Among all possible normalization schemes, we decide to subtract the column averages because we aim at identifying patterns within samples based on their correlation in bacterial composition. Indeed, the covariance matrix of the samples is proportional to $(X^p)^TX^p$, where $(X^p)^T$ is the transpose matrix. SVD on the matrix $X^p$ is thus equivalent to the principal component analysis (PCA) performed on the samples covariance matrix.

% Do NOT remove this, even if you are not including acknowledgments
\section*{Acknowledgments}
The authors acknowledge Chris Sander, Kevin Foster, Jonas Schluter, Carlos Carmona-Fontaine, Massimo Vergassola, Stefano Di Talia, Les Dethlefsen, Deb Bemis and two anonymous Reviewers for their insightful comments and help. S.B. acknowledges the GDRE 224 GREFI-MEFI CNRS-INdAM. This project was supported by NIH New Innovator Award (DP2OD008440) and Lucille Castori Center for Microbes Inflammation and Cancer grants to J.B.X.

%\section*{References}
% The bibtex filename
%\bibliographystyle{plos2009.bst}
\bibliography{PloSCBio_NoiseRef}

\pagebreak

\section*{Supplementary Information legends}

\subsection*{Supplementary Information Text} The Supplementary Information (SI) Text reports additional calculations, figures and details on: 1) model and relative stability analysis, 2) effect of random fluctuations and noise-induced dynamics and 3) Singular Value Decomposition.

\subsection*{Video S1} The video shows the stationary probability distributions $P_s$ as a function of the sensitive and tolerant densities for increasing noise value $D$, which ranges from $10^{-4}$ to $10^{-2}$. For visualization purposes, the noise value associated to each movie frame is displayed as an increasing bar in the top panel. 

\subsection*{Video S2} The video shows the time evolution of the two principal components  for the three subjects from \cite{Dethlefsen2010}. Empty circles represent untreated samples, asterisks represent samples during treatement 1 and filled circles represent represent samples during treatement 2. 

%\begin{table}[!ht]
%\caption{
%\bf{Table title}}
%\begin{tabular}{|c|c|c|}
%table information
%\end{tabular}
%\begin{flushleft}Table caption
%\end{flushleft}
%\label{tab:label}
% \end{table}

\end{document}